\documentclass[3p,times]{elsarticle}

\usepackage{ecrc}

\volume{00}

\firstpage{1}

\journalname{Nuclear Physics A}

\runauth{Author1 et al.}

\jid{NUPHA}

\jnltitlelogo{Nuclear Physics A}

\CopyrightLine{2012}{Published by Elsevier Ltd.}

\usepackage{amssymb}

\usepackage[figuresright]{rotating}

\newcommand{\La}{{\Lambda}}
\newcommand{\Si}{{\Sigma}}
\newcommand{\ccc}{\cdot\cdot\cdot}
\newcommand{\be}{\begin{eqnarray}}
\newcommand{\ee}{\end{eqnarray}}

\begin{document}

\begin{frontmatter}

\dochead{}

\title{Baryon-baryon interactions from chiral effective field theory}
\author{J. Haidenbauer}
\address{Institute for Advanced Simulation, Institut f{\"u}r Kernphysik (Theorie) and
J\"ulich Center for Hadron Physics, Forschungszentrum J{\"u}lich, D-52425 J{\"u}lich, Germany}

\begin{abstract}
Results from an ongoing study of baryon-baryon systems with strangeness 
$S=-1$ and $-2$ within chiral effective field theory are reported.
The investigations are based on the scheme proposed by Weinberg 
which has been applied rather successfully to the nucleon-nucleon 
interaction in the past.
Results for the hyperon-nucleon and hyperon-hyperon interactions obtained
to leading order are reviewed. Specifically, the issue of extrapolating
the binding energy of the $H$-dibaryon, extracted from recent lattice QCD simulations,
to the physical point is addressed. Furthermore, first results for the
hyperon-nucleon interaction at next-to-leading order are presented
and discussed.
\end{abstract}
\end{frontmatter}

\section{Introduction}
\label{intro}

Chiral effective field theory (EFT) as proposed in the pioneering works of 
Weinberg \cite{Wei90,Wei91} is a powerful tool for the derivation of nuclear forces.
In this scheme there is an underlying power counting which allows to improve calculations 
systematically by going to higher orders in a perturbative expansion. 
In addition, it is possible to derive two- and corresponding three-nucleon forces as well 
as external current operators in a consistent way. 
Over the last decade or so it has been demonstrated that 
the nucleon-nucleon ($NN$) interaction can be described to a high precision 
within the chiral EFT approach \cite{Entem:2003ft,Epe05}. Following the original
suggestion of Weinberg, in these works the power counting is applied to the $NN$ 
potential rather than to the reaction amplitude. The latter is then obtained from 
solving a regularized Lippmann-Schwinger equation for the derived interaction potential. 
The $NN$ potential contains pion-exchanges and a series of contact interactions 
with an increasing number of derivatives to parameterize the shorter ranged part 
of the $NN$ force.
For reviews we refer the reader 
to Refs.~\cite{Bed02,Epelbaum:2005pn,Epelbaum:2008ga}. 
 
In the present contribution I focus on recent investigations by the
groups in Bonn-J\"ulich and Munich on the baryon-baryon interaction involving strange 
baryons, performed within chiral EFT 
\cite{Polinder:2006zh,Polinder:2007mp,Hai10a,YNNL1,YNNL2}. 
In these works the same scheme as applied in Ref.~\cite{Epe05} to the $NN$ 
interaction is adopted. First I discuss the application to the strangeness $S=-1$
sector ($\Lambda N$, $\Sigma N$). Here the extension of our study \cite{Polinder:2006zh} 
to next-to-leading order (NLO) is in progress \cite{YNNL2} and a first glimpse on 
the (still preliminary) 
achieved results for the $\Lambda N$ and $\Sigma N$ interactions will be given.
Then I report results of a study on the strangeness $S=-2$ sector, i.e. for the 
$\Lambda\Lambda$, $\Sigma\Sigma$, and cascade-nucleon ($\Xi N$) interactions.
Predictions obtained at leading order (LO) \cite{Polinder:2007mp} are reviewed 
and implications
for the $H$-dibaryon are discussed, based on our framework, in the light of 
recent lattice QCD calculations where evidence for the existence of such a 
state was found. 
 
At LO in the power counting, as considered in the 
aforementioned investigations \cite{Polinder:2006zh,Polinder:2007mp,Hai10a}, 
the baryon-baryon 
potentials involving strange baryons consist of four-baryon contact terms without 
derivatives and of one-pseudoscalar-meson exchanges, analogous to the $NN$ 
potential of \cite{Epe05}. 
The potentials are derived using constraints from ${\rm SU(3)}$ flavor symmetry. 
At NLO one gets contributions from 
two-pseudoscalar-meson exchange diagrams and from four-baryon contact terms 
with two derivatives \cite{Epe05}. 

The paper is structured as follows: 
In Sect. 2 a short overview of the chiral EFT approach is provided. 
In Sect. 3 results for the $\Lambda N$- and $\Sigma N$ interactions 
obtained to NLO are presented. 
In Sect. 4 results for the the $S=-2$ ($\Lambda \Lambda$, $\Xi N$ $\Si\Si$) 
systems are briefly reviewed and connection is made with lattice QCD results 
for the $H$-dibaryon case. 
The paper ends with a short Summary.  

\section{Formalism}
\label{sec:2}
The derivation of the chiral baryon-baryon potentials for the strangeness sector 
at LO using the Weinberg power counting is outlined in Refs. 
\cite{Polinder:2006zh,Hai10a,Haidenbauer:2007ra}. Details for the NLO case will 
be presented in a forthcoming paper \cite{YNNL2}, see also \cite{YNNL1,Petschauer}. 
The LO potential consists of four-baryon contact terms without derivatives and of 
one-pseudoscalar-meson exchanges while at NLO contact terms with two derivatives 
arise, together with contributions from (irreducible) two-pseudoscalar-meson exchanges.
 
The spin- and momentum structure of the potentials resulting from the contact terms to LO is given by 
\begin{eqnarray}
V^{(0)}_{BB\to BB} &=& C_{S;\,BB\to BB} + C_{T;\,BB\to BB}
(\mbox{\boldmath $\sigma$}_1\cdot\mbox{\boldmath $\sigma$}_2)
\end{eqnarray}
in the notation of \cite{Epe05} where the $C_{i;\,BB\to BB}$'s are so-called low-energy 
coefficients (LECs) that need to be determined by a fit to data. 
Due to the imposed ${\rm SU(3)}_{\rm f}$ constraints there are only five independent 
LECs for the $NN$ and the $YN$ sectors together,
as described in Ref.~\cite{Polinder:2006zh} where also the relations between the
various $C_{i;\,BB\to BB}$'s are given. 
A sixth LEC is, however, present in the strangeness $S=-2$ channels with isospin $I=0$. 
 
In next-to-leading order one gets the following spin- and momentum structure:
\begin{eqnarray}
V^{(2)}_{BB\to BB} &=& C_1 {\bf q}^{\,2}+ C_2 {\bf k}^{\,2} + (C_3 {\bf q}^{\,2}+ C_4 {\bf k}^{\,2})
(\mbox{\boldmath $\sigma$}_1\cdot\mbox{\boldmath $\sigma$}_2) 
+ i C_5 (\mbox{\boldmath $\sigma$}_1+\mbox{\boldmath $\sigma$}_2)\cdot ({\bf q} \times {\bf k}) \nonumber \\
&+& C_6 ({\bf q} \cdot \mbox{\boldmath $\sigma$}_1) ({\bf q} \cdot \mbox{\boldmath $\sigma$}_2)
+ C_7 ({\bf k} \cdot \mbox{\boldmath $\sigma$}_1) ({\bf k} \cdot \mbox{\boldmath $\sigma$}_2)
+ i C_8 (\mbox{\boldmath $\sigma$}_1-\mbox{\boldmath $\sigma$}_2)\cdot ({\bf q} \times {\bf k}). 
\end{eqnarray}
The transferred 
and average momentum, ${\bf q}$ and ${\bf k}$, are defined in terms of the final and initial 
center-of-mass (c.m.) momenta of the baryons, ${\bf p}'$ and ${\bf p}$, as 
${\bf q}={\bf p}'-{\bf p}$ and ${\bf k}=({\bf p}'+{\bf p})/2$. 
The $C_i$'s (actually $C_{i;\,BB\to BB}$'s) are additional LECs. Performing a partial wave 
projection and imposing again ${\rm SU(3)}_{\rm f}$ symmetry one finds that in case of the 
$YN$ interaction there are eight new LECs entering the $S$-waves and $S$-$D$ transitions, 
respectively, and ten coefficents in the $P$-waves. There are further (four) LECs 
that contribute only to the $S=-2$ system. 

The spin-space part of the one-pseudoscalar-meson-exchange potential is similar to the 
static one-pion-exchange potential (recoil and relativistic corrections give 
higher order contributions) and follows from the ${\rm SU(3)}_{\rm f}$ invariant 
pseudoscalar-meson--baryon interaction Lagrangian with the appropriate symmetries as 
discussed in \cite{Polinder:2006zh}:
\begin{eqnarray}
V^{OBE}&=&-f_{B_1B_1'P}f_{B_2B_2'P}\frac{\left(\mbox{\boldmath $\sigma$}_1\cdot{\bf q}\right)\left(\mbox{\boldmath $\sigma$}_2\cdot{\bf q}\right)}{{\bf q}^2+M^2_P}\ .
\label{POT}
\end{eqnarray}
Here, $M_P$ is the mass of the exchanged pseudoscalar meson. 
The coupling constants $f_{BB'P}$ at the various baryon-baryon-meson vertices are fixed
by the imposed ${\rm SU(3)}$ constraints and tabulated, e.g., in \cite{Polinder:2006zh}. 
They can be expressed in terms of 
$f\equiv g_A/2F_\pi\equiv f_{NN\pi}$ ($g_A= 1.26$, $F_\pi =  92.4$ MeV) and $\alpha$, the so-called  
$F/(F+D)$-ratio, for which we adopted the ${\rm SU(6)}$ value ($\alpha=0.4$).
Note that we use the physical masses of the exchanged pseudoscalar mesons. 
Thus, the explicit ${\rm SU(3)}$ breaking reflected in the mass splitting between the 
pseudoscalar mesons is taken into account. 
The $\eta$ meson was identified with the octet $\eta$ ($\eta_8$) and its physical 
mass was used.
The two-pseudoscalar-meson-exchange potential can be found in Refs.~\cite{YNNL1,YNNL2}. 

The reaction amplitudes are obtained from the solution of a coupled-channels Lippmann-Schwinger (LS) 
equation for the interaction potentials: 
\begin{eqnarray}
&&T_{\rho''\rho'}^{\nu''\nu',J}(p'',p';\sqrt{s})=V_{\rho''\rho'}^{\nu''\nu',J}(p'',p')+
\sum_{\rho,\nu}\int_0^\infty \frac{dpp^2}{(2\pi)^3} \, V_{\rho''\rho}^{\nu''\nu,J}(p'',p)
\frac{2\mu_{\nu}}{q_{\nu}^2-p^2+i\eta}T_{\rho\rho'}^{\nu\nu',J}(p,p';\sqrt{s})\ .
\label{LS} 
\end{eqnarray}

The label $\nu$ indicates the particle channels and the label $\rho$ the partial wave. $\mu_\nu$ 
is the pertinent reduced mass. The on-shell momentum in the intermediate state, $q_{\nu}$, is 
defined by $\sqrt{s}=\sqrt{m^2_{B_{1,\nu}}+q_{\nu}^2}+\sqrt{m^2_{B_{2,\nu}}+q_{\nu}^2}$. 
Relativistic kinematics is used for relating the laboratory energy $T_{{\rm lab}}$ of the hyperons 
to the c.m. momentum.

We solve the LS equation in the particle basis, in order to incorporate the correct physical
thresholds. Depending on the specific values of strangeness and charge up to six baryon-baryon
channels can couple. For the $S=-1$ sector where a comparison with scattering data is 
possible the Coulomb interaction is taken into account appropriately via the Vincent-Phatak
method \cite{VP}. 
The potentials in the LS 
equation are cut off with a regulator function, $\exp\left[-\left(p'^4+p^4\right)/\Lambda^4\right]$, 
in order to remove high-energy components of the baryon and pseudoscalar meson fields \cite{Epe05}.
We consider cut-off values in the range 500, ..., 700 MeV, similar to what was used for  
chiral $NN$ potentials \cite{Epe05}.

\begin{figure}[t]
\begin{center}
\includegraphics[height=56mm]{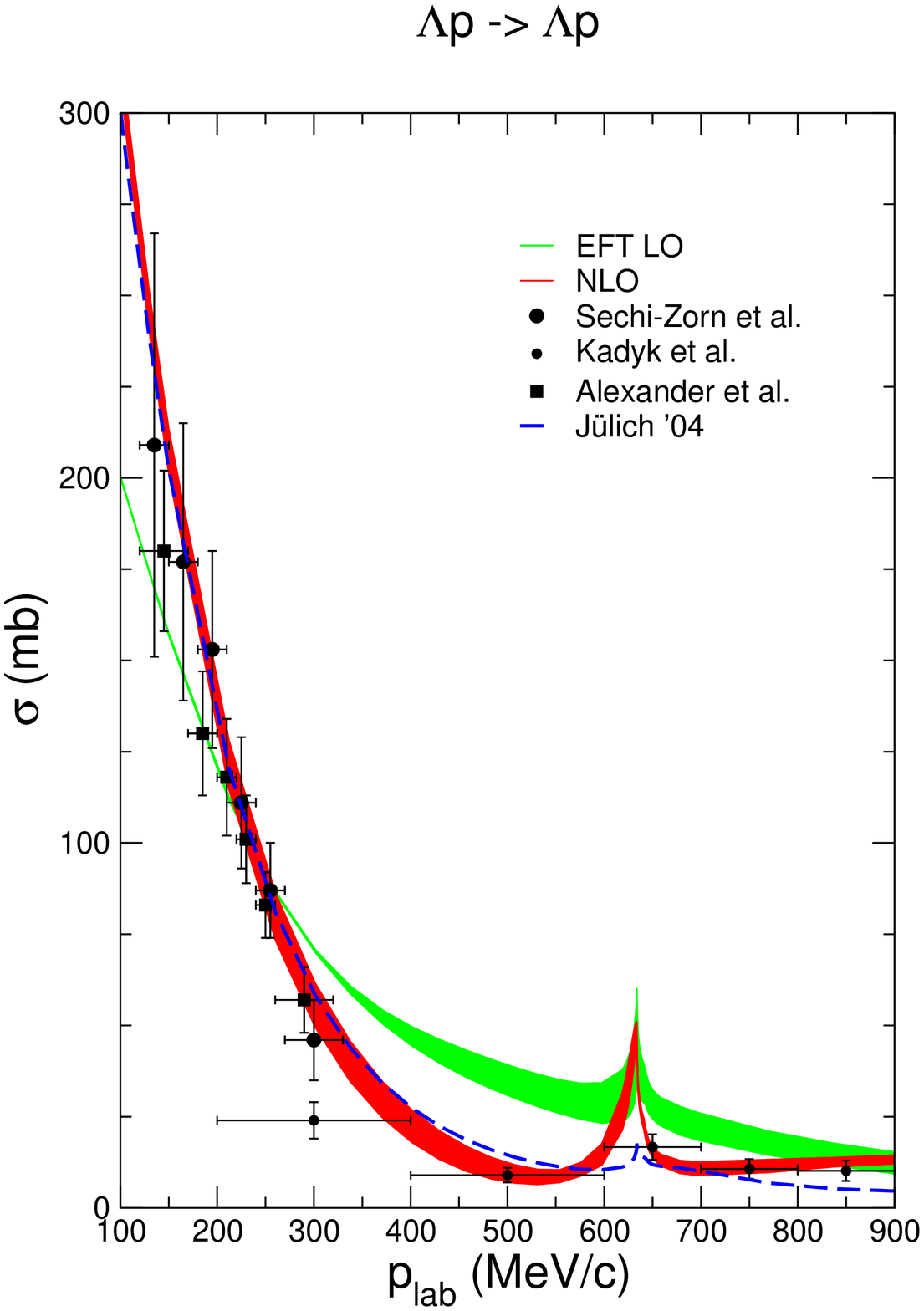}
\includegraphics[height=56mm]{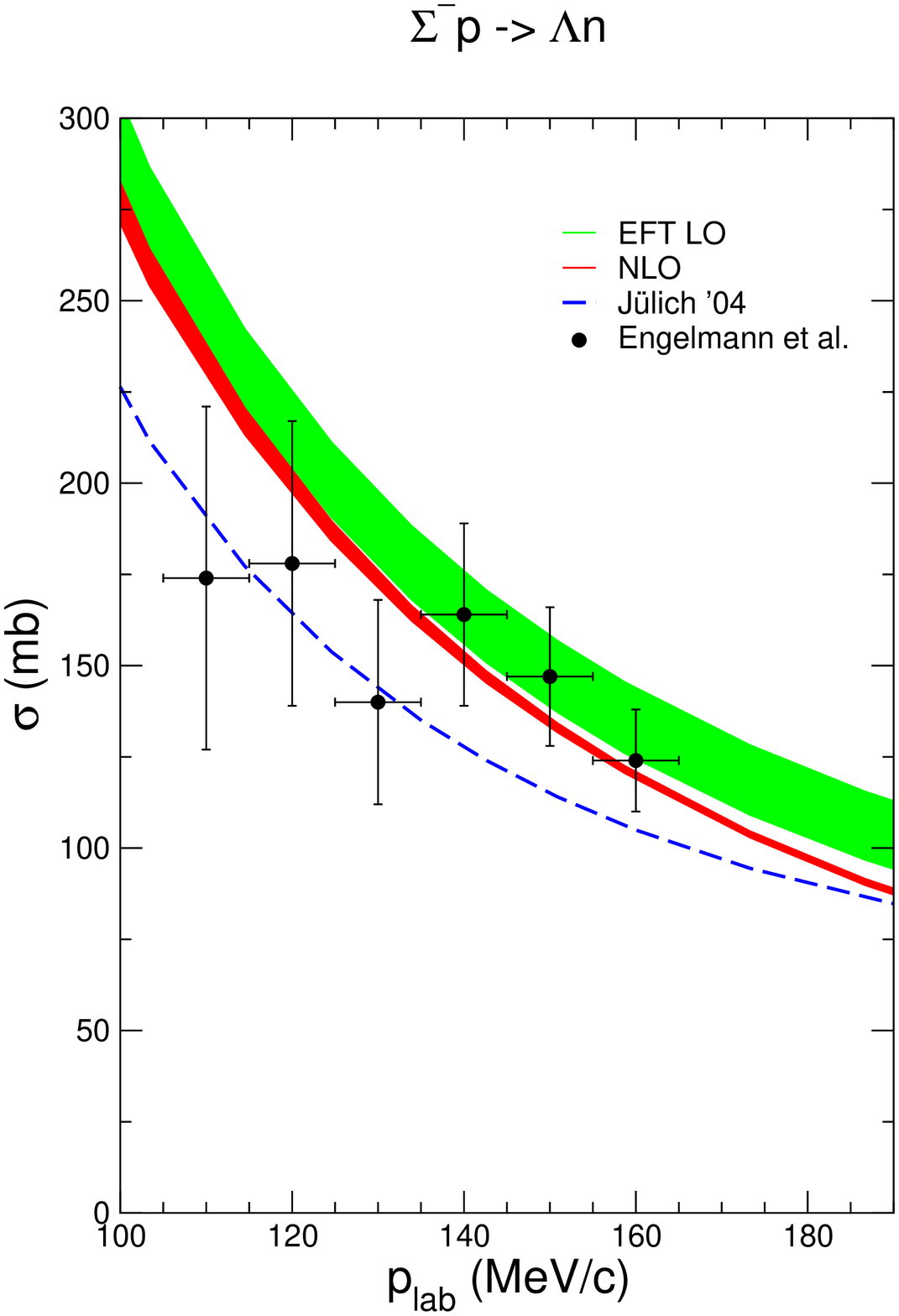}
\includegraphics[height=56mm]{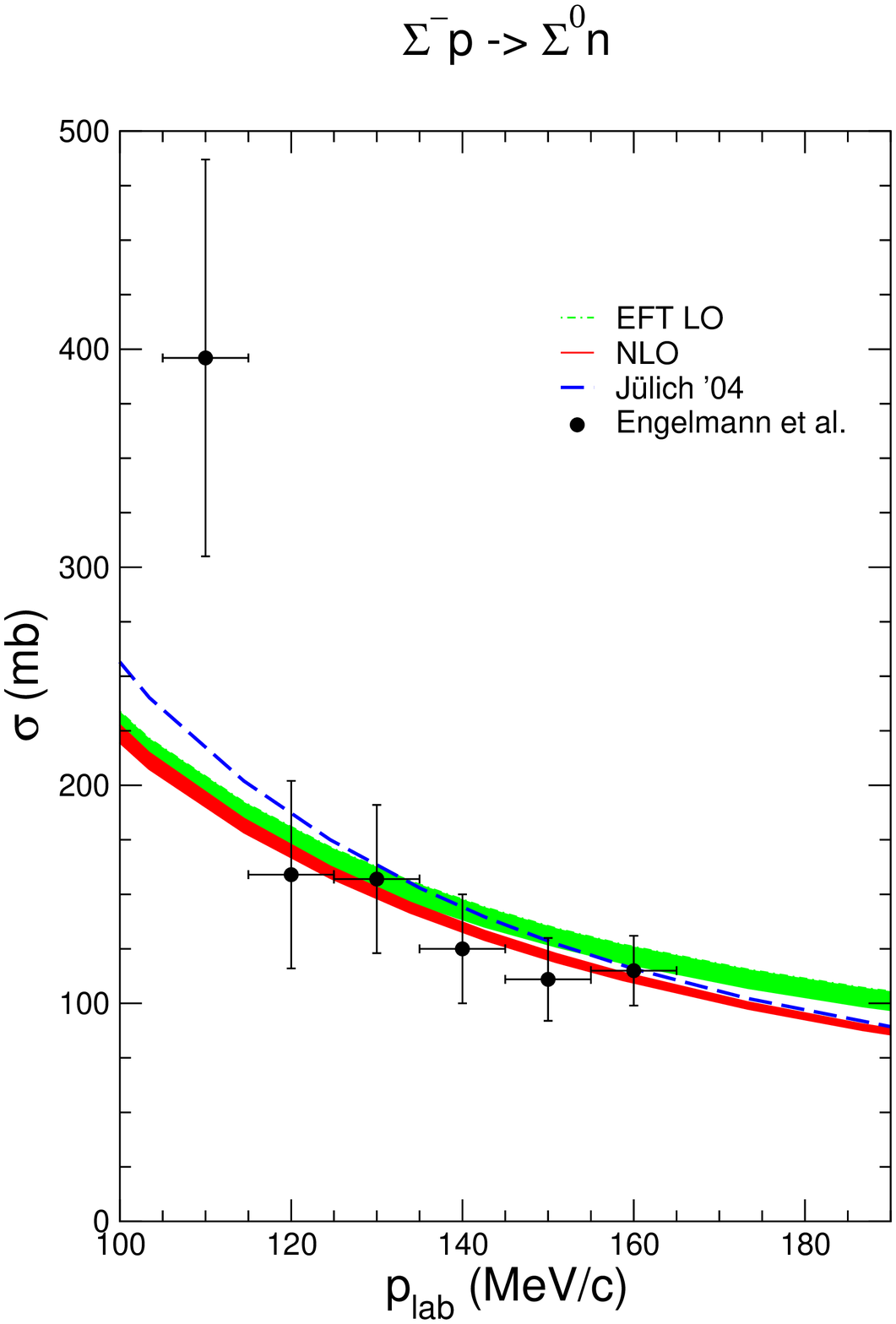}

\includegraphics[height=56mm]{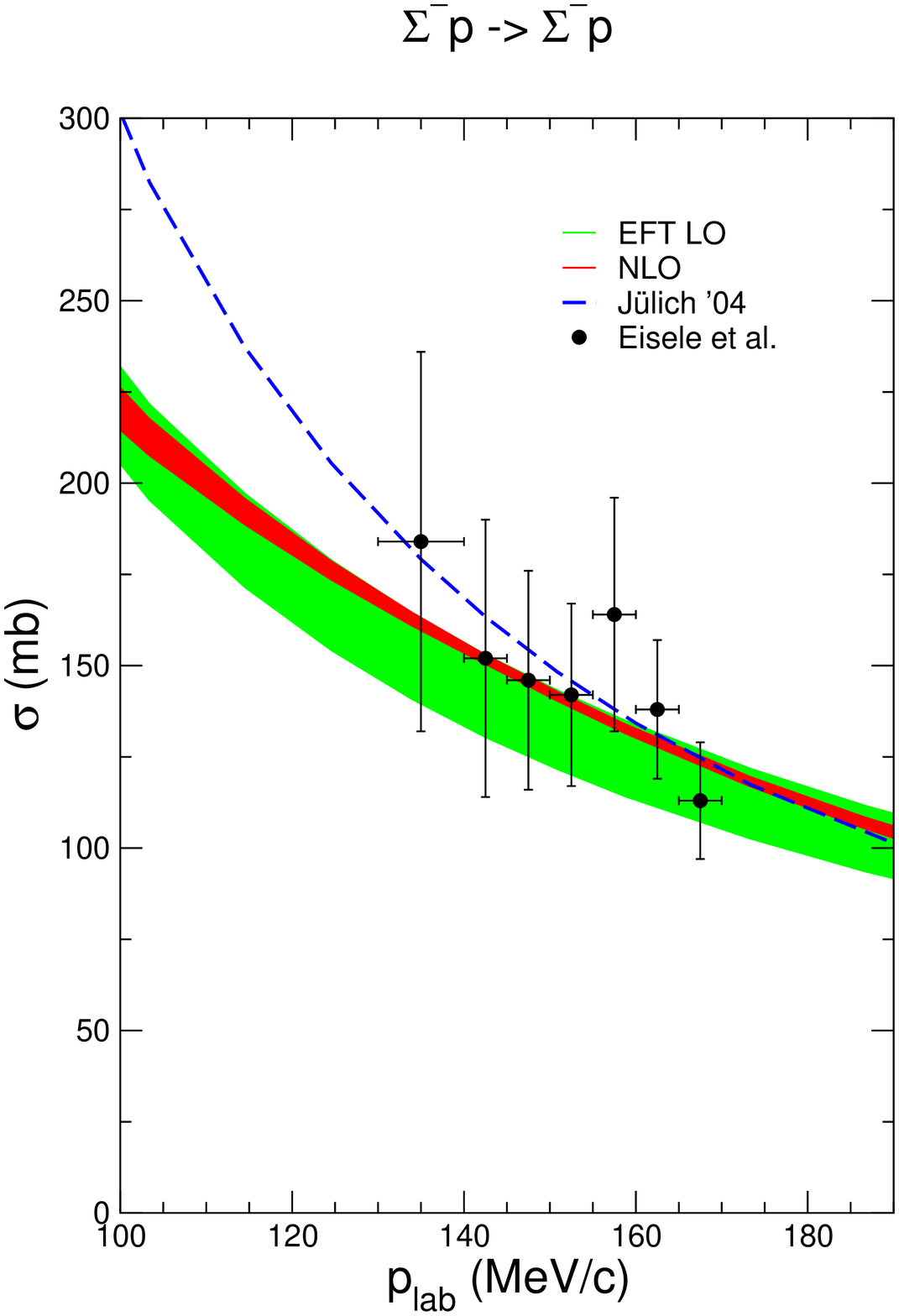}
\includegraphics[height=56mm]{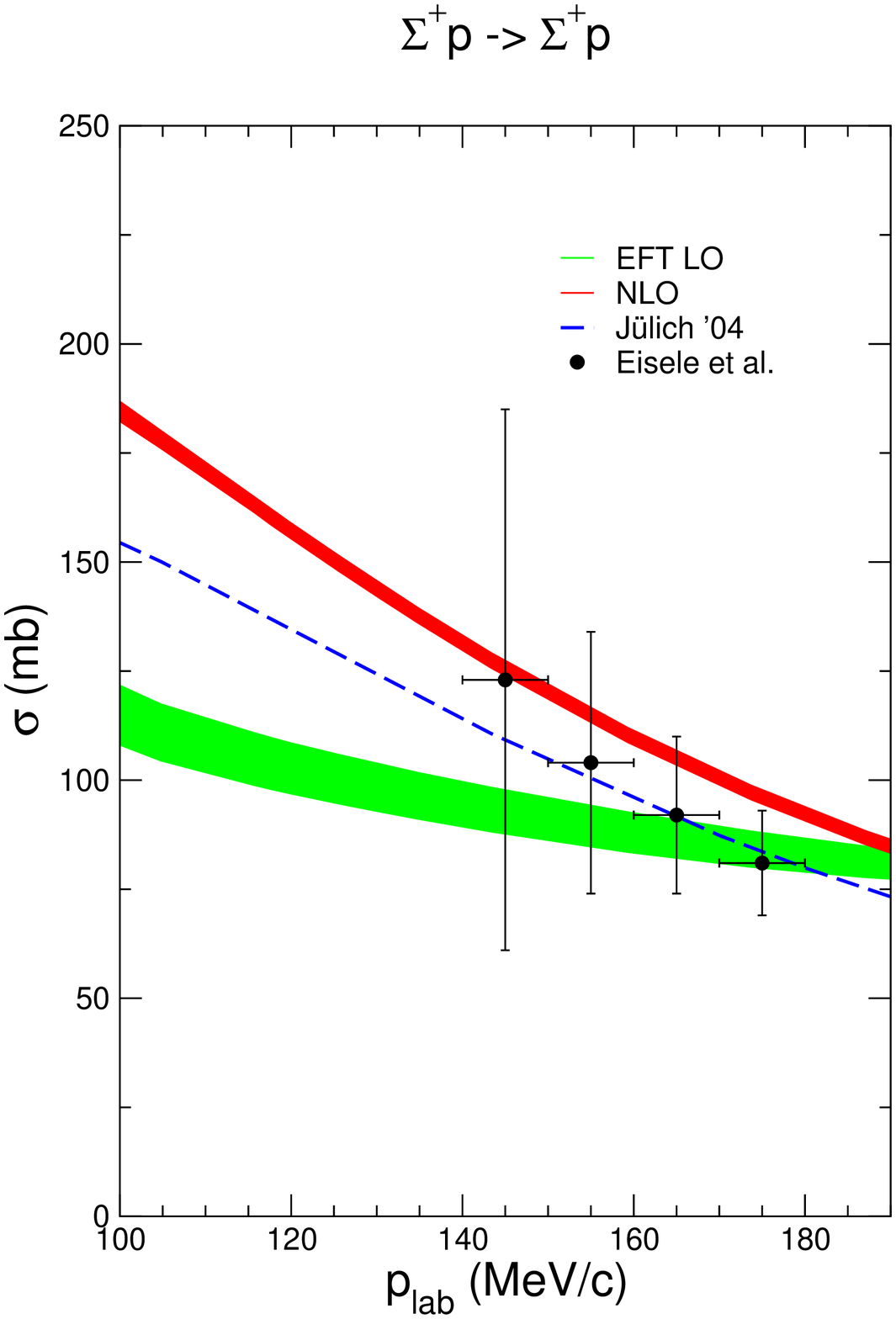}
\caption{Total cross sections for $\Lambda p \to \Lambda p$, $\Sigma^-p \to \Lambda n$,
$\Sigma^-p \to \Sigma^0 n$, $\Sigma^-p \to \Sigma^- p$ and 
$\Sigma^+p \to \Sigma^+ p$ as a function of $p_{lab}$.
The green (grey) band shows the chiral EFT results to LO for variations of the cut-off
in the range $\Lambda =$ 550$\ldots$700~MeV, while the red (black) band are
results to NLO for $\Lambda =$ 500$\ldots$700~MeV. The dashed curve is the result of 
the J{\"u}lich '04 \cite{Hai05} meson-exchange potential. 
}
\label{fig:1}
\end{center}
\end{figure}

\section{Results for the strangeness S=-1 sector}

The imposed ${\rm SU(3)}$ flavor symmetry implies that at LO five independent LECs 
contribute to the $YN$ interaction \cite{Polinder:2006zh}. 
These five contact terms were determined in \cite{Polinder:2006zh} by a fit to 
the $YN$ scattering data. Already in that scenario a fairly reasonable description 
of the 35 low-energy $YN$ scattering data could be achieved for cutoff 
values $\Lambda=550,...,700$ MeV and for natural values of the LECs. 
At NLO there are eight new contact terms contributing to the $S$-waves and
the $^3S_1$-$^3D_1$ transition, and ten in
the $P$-waves. Once again the corresponding LECs were fixed by fitting to the data.
The results obtained at NLO are presented in Fig.~\ref{fig:1} (black (red) bands), 
together with those at LO (grey (green) bands). The bands represent the variation 
of the cross sections based on chiral EFT within the considered cutoff region. 
For comparison also results for the 
J{\"u}lich '04 \cite{Hai05} meson-exchange models are shown (dashed line).

Obviously, and as expected, the energy dependence exhibited by the data can be significantly 
better reproduced within our NLO calculation. This concerns in particular the $\Sigma^+p$
channel. But also for $\Lambda p$ the NLO results are now well in line with the data even 
up to the $\Sigma N$ threshold. Furthermore, one can see that the dependence on the cutoff
mass is strongly reduced in the NLO case. 

Note that in case of LO as well as at NLO no ${\rm SU(3)}_{\rm f}$ constraints 
from the $NN$ sector were imposed in the fitting procedure. 
The leading order ${\rm SU(3)}_{\rm f}$ breaking in the one-boson
exchange diagrams (coupling constants) is ignored. 

Besides an excellent description of the $YN$ data the chiral EFT
interaction also yields a correctly bound hypertriton, see Table~\ref{tabb}.
Indeed this binding energy had to be included in the fitting procedure
because otherwise it would have not been possible to fix the relative strength 
of the (S-wave) singlet- and triplet contributions to the $\Lambda p$ interaction. 
Table~\ref{tabb} lists also results for two meson-exchange potentials, namely
of the J\"ulich '04 model \cite{Hai05} and the Nijmegen NSC97f potential \cite{Rij99},
which both reproduce the hypertriton binding energy correctly. 
Obviously, the scattering lengths predicted at NLO are
larger than those obtained at LO and now similar to the values of the meson-exchange
potentials. The $\Si^+ p$ scattering length in the $^3S_1$ partial wave is positive,
as it was already the case for our LO potential, indicating a repulsive interaction 
in this channel. 

\begin{table}[h]
\begin{center}
\begin{tabular}{|c|r|r|rr|r|}
\hline
& \ EFT LO \ & \ EFT NLO \ & \ J\"ulich '04 \cite{Hai05} \ & \ NSC97f \cite{Rij99} \ & \ experiment \ \\
\hline
${\Lambda}$ [MeV] & 550 {$\ccc$} 700 & 500 {$\ccc$} 700 & & & \\
\hline
$a^{\La p}_s$ & $-1.90$ {$\ccc$} $-1.91$ & $-2.88$ {$\ccc$} $-2.89$ & $-2.56$ & $-2.51$ & $-1.8^{+2.3}_{-4.2}$\\
$a^{\La p}_t$ & $-1.22$ {$\ccc$} $-1.23$ & $-1.59$ {$\ccc$} $-1.61$ & $-1.66$ & $-1.75$ & $-1.6^{+1.1}_{-0.8}$\\
\hline
$a^{\Si^+ p}_s$ & $-2.24$ {$\ccc$} $-2.36$ & $-3.90$ {$\ccc$} $-3.83$ & $-4.71$ & $-4.35$ &\\
$a^{\Si^+ p}_t$ & $0.70$ {$\ccc$}  $0.60$  & $ 0.51$ {$\ccc$} $0.47$ & $0.29$ & $-0.25$ &\\
\hline
\hline
$(^3_\La \rm H)$ $E_B$ & $-2.34$ {$\ccc$} $-2.36$ & $-2.31$ {$\ccc$} $-2.34$ & $-2.27$ & $-2.30$ & -2.354(50)\\
\hline
\end{tabular}
\caption{The $YN$ singlet (s) and triplet (t) scattering lengths (in fm) 
  and the hypertriton binding energy, $E_B$ (in MeV). 
  }
\label{tabb}
\end{center}
\end{table}

Calculations for the four-body hypernuclei ${}^4_\Lambda {\rm H}$ and ${}^4_\Lambda {\rm He}$ 
based on those interactions are reported in Ref.~\cite{Nog12}.

\section{Results for the strangeness $S=-2$ sector}

\begin{figure}[t]
\centering
\resizebox{1.00\textwidth}{!}{%
  \includegraphics*[2.0cm,17.0cm][15.65cm,26.8cm]{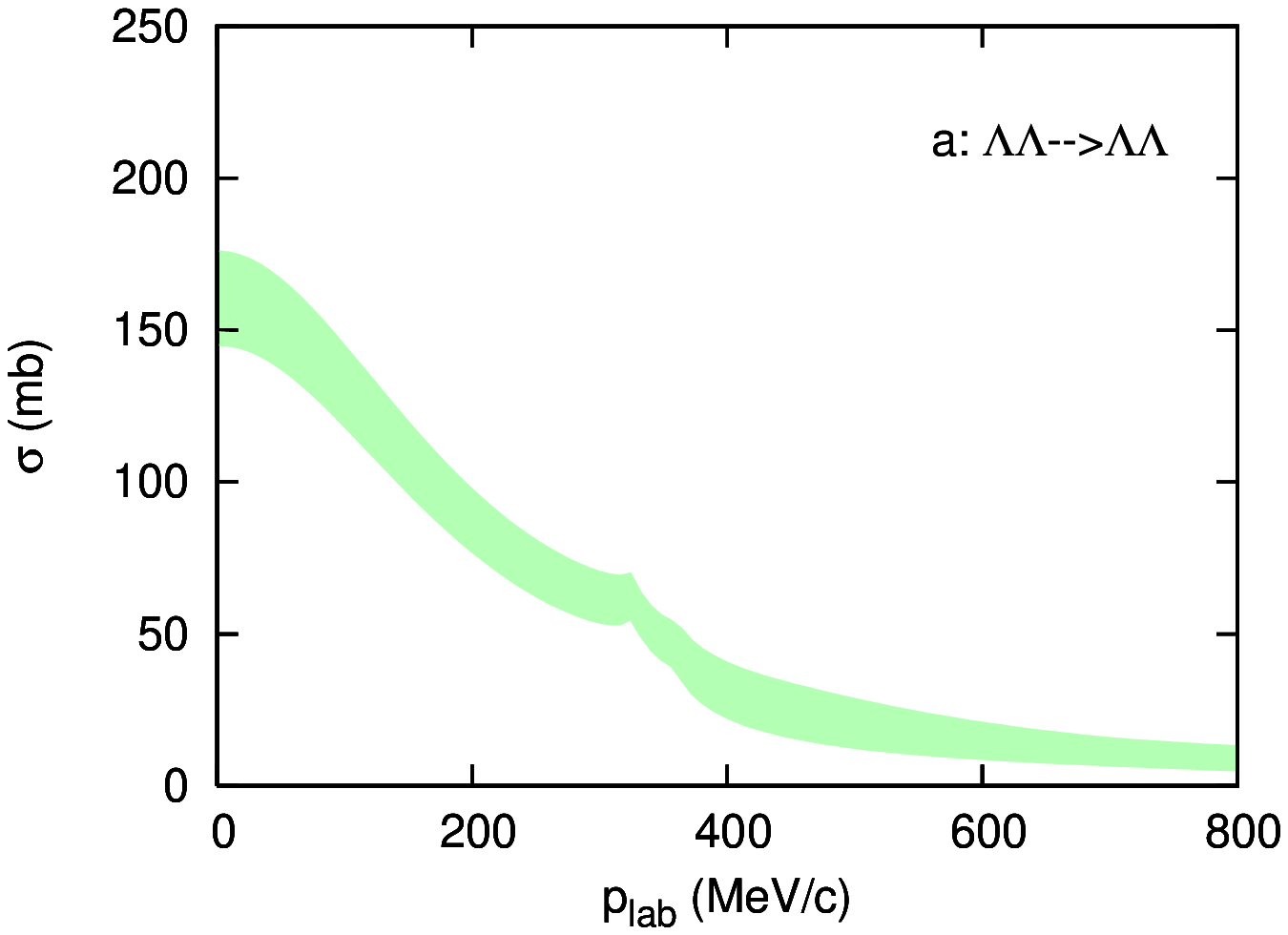}
  \includegraphics*[2.0cm,17.0cm][15.65cm,26.8cm]{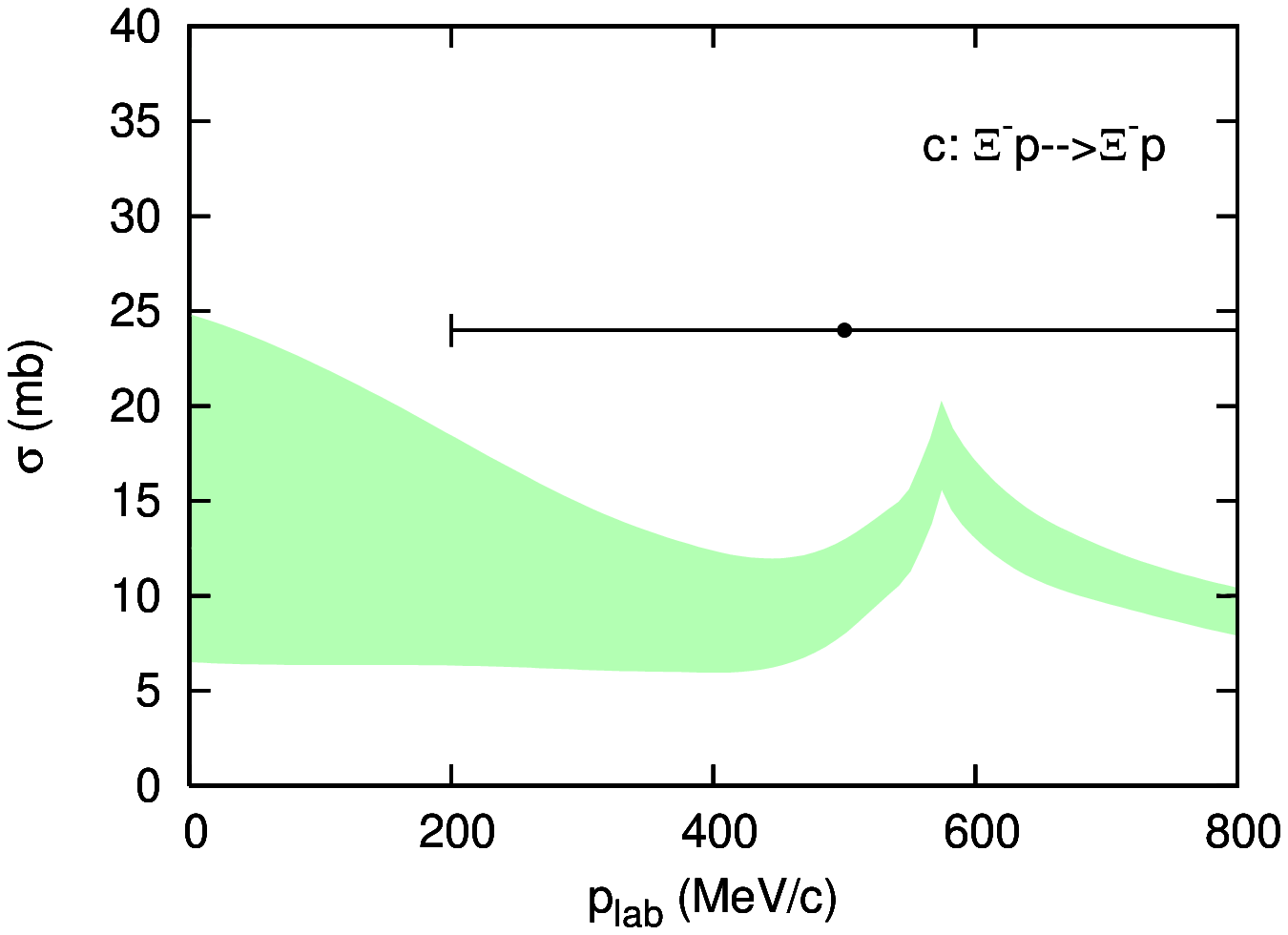}
  \includegraphics*[2.0cm,17.0cm][15.65cm,26.8cm]{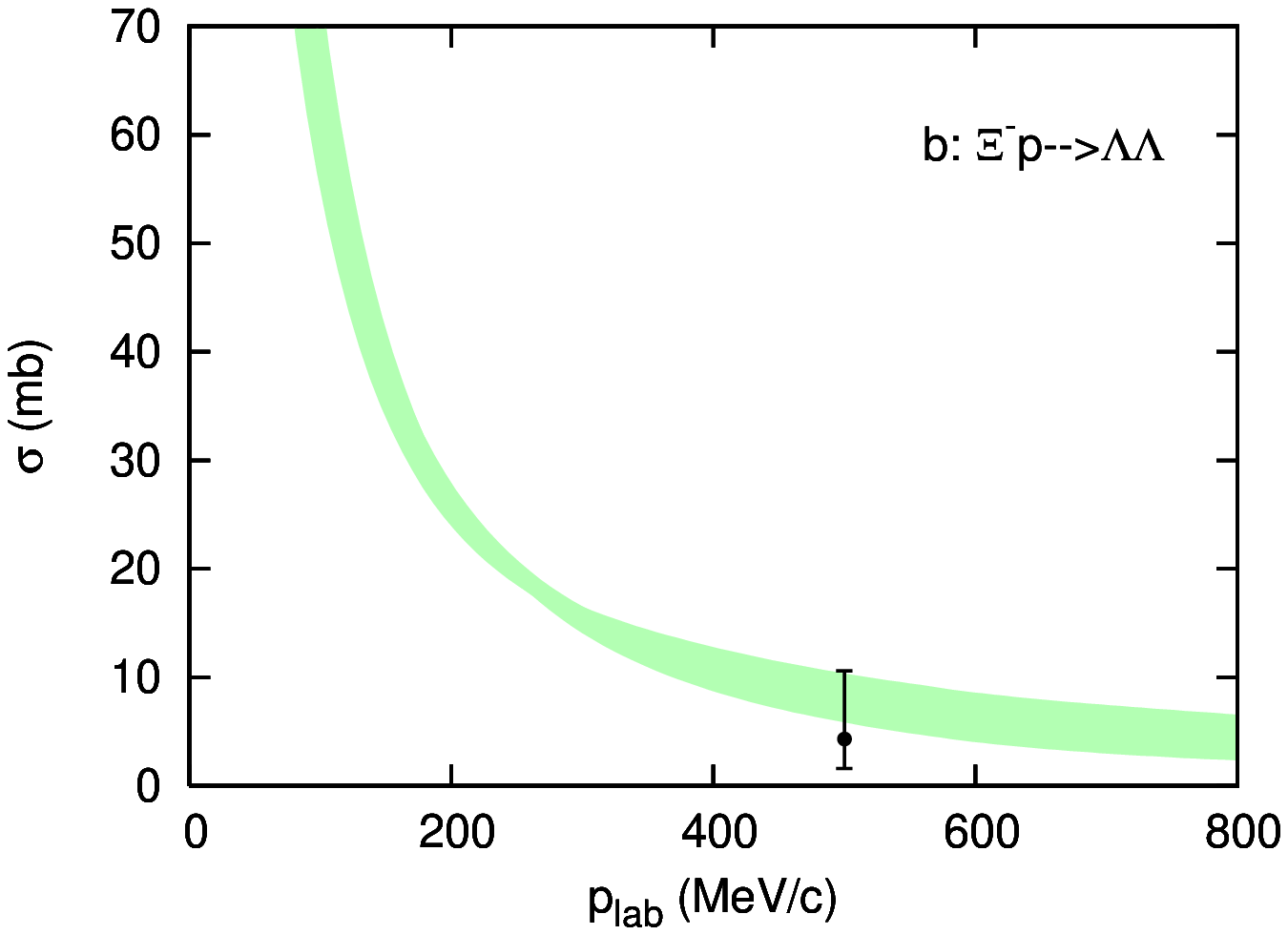}
}
\caption{$YY$ and $\Xi N$ integrated cross sections as a function of $p_{\rm lab}$.
The band shows the chiral EFT result at LO for variations of the cutoff $\Lambda$ as 
discussed in the text. Data are from Ref.~\cite{Ahn:2005jz}.
}
\label{fig:2}
\end{figure}

In this section I review results obtained for the $S=-2$ sector, specifically
for the coupled $\La\La - \Xi N - \Si\Si$ system, within chiral EFT at LO
\cite{Polinder:2007mp,Haidenbauer11,Haidenbauer12}. 
As mentioned above, at LO one additional LEC occurs in this specific 
channel with $I=0$ which, in principle, should be determined from 
experimental information available for this sector. However, the scarce 
data ($\Xi^-p \to \Xi^-p$ and $\Xi^-p \to \Lambda\Lambda$ 
cross sections \cite{Ahn:2005jz}) are afflicted with large uncertainties 
and, thus, do not allow to establish reliably its value as found by us
\cite{Polinder:2007mp}. 
Some results for strangeness $S=-2$ published in \cite{Polinder:2007mp} are 
reproduced here in Fig. \ref{fig:2}. 
As before the band reflects the dependence of the results on variations of
the cutoff $\Lambda$. The cutoff was varied between 550 and 700 MeV (like in case
of the LO $YN$ potential) and under the constraint that the $\Lambda\Lambda$ $^1S_0$ 
scattering
length remains practically unchanged \cite{Polinder:2007mp}. As reference we
have taken the result for $\Lambda = 600$ MeV and with the value of the 
additional LEC fixed in such a way that $C_{\La\La\to\La\La} = 0$. 
The scattering length turned out to be $a_{s}^{\Lambda\Lambda} = -1.52$ fm \cite{Polinder:2007mp}. 
Analyses of the measured binding energy of the double-strange hypernucleus
${}^{\;\;\;6}_{\Lambda\Lambda}{\rm He}$ \cite{Takahashi:2001nm} suggest that
the $\La\La$ scattering length could be in the range of
$-1.3$ to $-0.7$ fm \cite{Gal,Rijken,Fujiwara}.
A first determination of the scattering length utilizing data on the
$\Lambda\Lambda$ invariant mass from the reaction
$^{12}C(K^-,K^+\Lambda\Lambda X)$ \cite{Yoon} led to the result
$a_s^{\Lambda\Lambda}=-1.2\pm 0.6$ fm \cite{Ashot}.


One particular interesting aspect of the coupled $\La\La - \Xi N - \Si\Si$ 
system is the $H$-dibaryon, 
a deeply bound 6-quark state with $J=0$ predicted by Jaffe from the bag model 
\cite{Jaffe:1976yi}, that should occur in this channel.
So far none of the experimental searches for the $H$-dibaryon let to 
convincing signals \cite{Yoon}. 
However, recently evidence for a bound $H$-dibaryon was claimed based on
lattice QCD calculations \cite{Beane,Inoue,Beane11a,Inoue11a}. Extrapolations 
of those computations, performed for $m_\pi \gtrsim 400$ MeV, to the physical pion 
mass suggest that the $H$-dibaryon could be either loosely bound or move into
the continuum \cite{Beane11,Shanahan11}.

Unfortunately, because the mentioned additional contact term cannot be 
reliably fixed from the data in the $S=-2$ sector, no immediate predictions 
can be made for the $H$-dibaryon within chiral EFT. 
However, the framework can be used as a tool to study
the dependence of its conjectured binding energy on the masses of the 
involved hadrons and, thus, allows for an alternative
extrapolation of the results, obtained in lattice QCD calculations 
at unphysical meson- and baryon masses, down to the physical point. 
In particular one can fine-tune the ``free'' LEC to produce a bound $H$ with 
a given binding energy for meson- and baryon masses corresponding to the
lattice simulations, and then study its properties \cite{Haidenbauer11,Haidenbauer12} 
for physical masses. Note that at LO the LECs do not depend on the meson masses
(strictly speaking, on the quark masses). 
Variations of the meson masses enter only in the potential in Eq.~(\ref{POT}), those
of the baryon masses (via $\mu_\nu$) in the LS equation~(\ref{LS}). 

\begin{figure}[t!]
\centering
\includegraphics[width=0.305\textwidth,keepaspectratio,angle=-90]{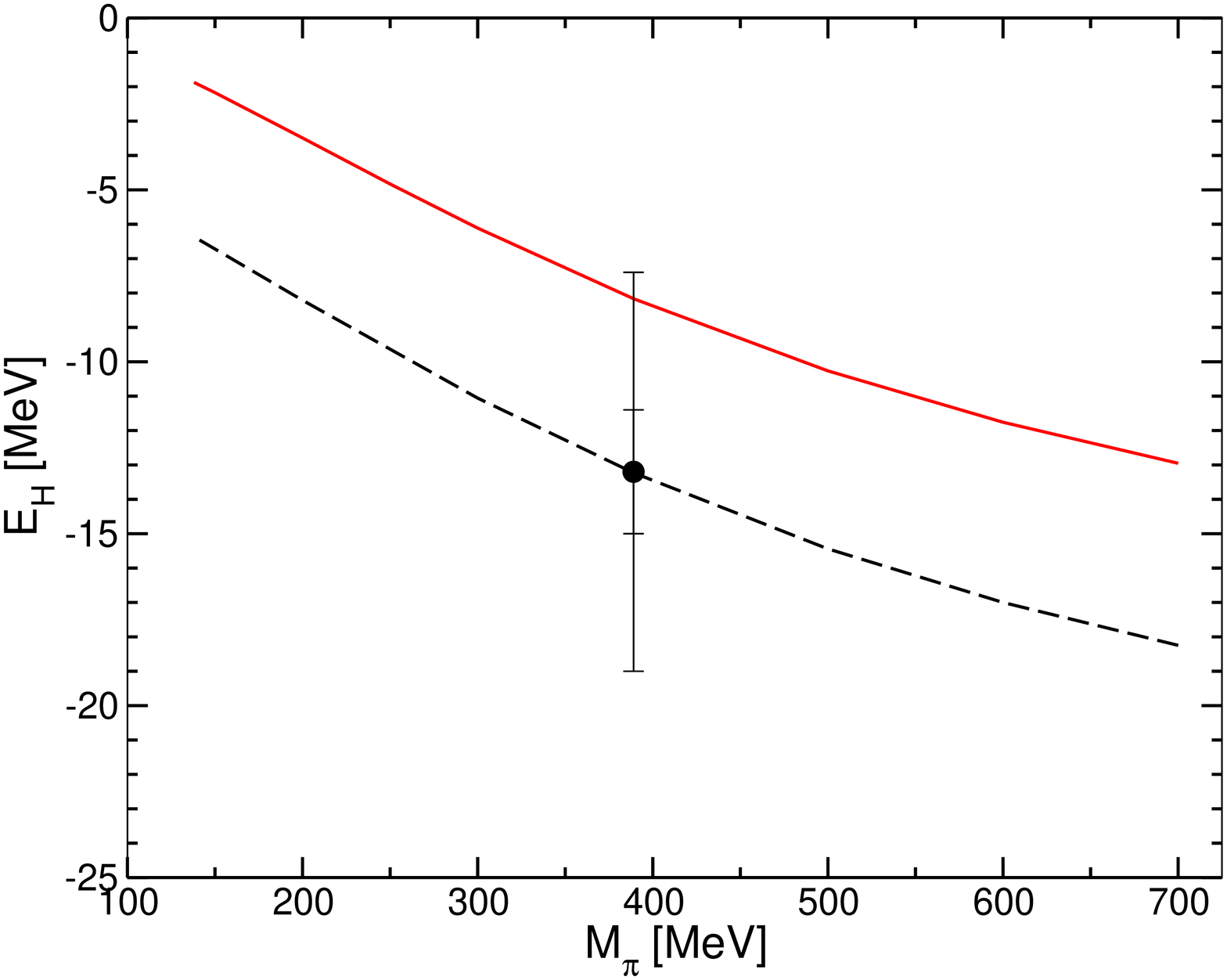}
\includegraphics[width=0.305\textwidth,keepaspectratio,angle=-90]{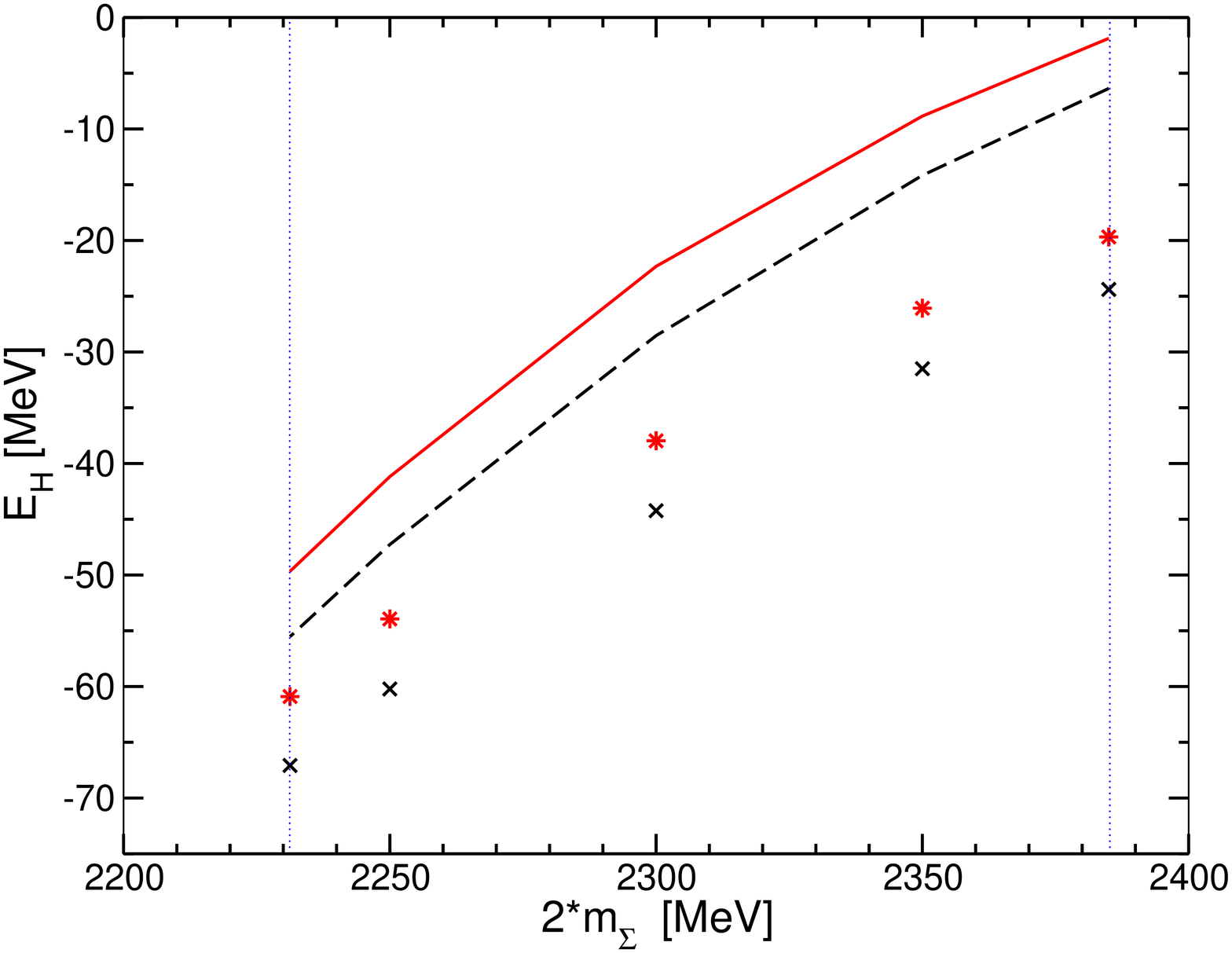}
\caption{Dependence of the binding energy of a $H$-dibaryon on the pion mass 
$M_\pi$ (left) and on the $\Sigma$ mass $m_\Sigma$ (right).
The solid curve correspond to the case where the LEC is fixed such that $E_H = -1.87\,$MeV 
for physical masses while for the dashed curve it is fixed to yield $E_H = -13.2\,$MeV for 
$M_\pi = 389$ MeV. 
The asterisks and crosses represent results where, besides the 
variation of $m_\Sigma$, $m_\Xi+m_N = 2 m_\La$ is assumed so that the $\Xi N$ threshold 
coincides with that of the $\La\La$ channel. 
The vertical (dotted) lines indicate the physical $\La\La$ and $\Si\Si$ thresholds. 
The circle indicates the lattice result of the NPLQCD Collaboration~\cite{Beane11a}.
}
\label{fig:mpi}
\end{figure}

Let me first consider individual variations of the masses of the involved particles. 
To begin with I examine the dependence of the $H$ binding energy on the pion mass $M_\pi$ 
and keep all other (meson and baryon) masses at their physical value. 
Corresponding results are displayed in Fig.~\ref{fig:mpi} (left). 
Adjusting the LEC so that a $H$ binding energy of 13.2 MeV is predicted for
$M_\pi=389$ MeV, corresponding to the result published by 
NPLQCD~\cite{Beane11a}, yields the dashed curve. 
The solid curve corresponds to a $H$-dibaryon that is bound by 1.87 MeV at the
physical point, i.e. with the same binding wave number (0.23161 fm$^{-1}$) as 
the deuteron in the $NN$ case. Enlarging the pion mass to around 400 MeV for 
the latter scenario (i.e. to values in an order that corresponds to the NPLQCD 
calculation \cite{Beane}) increases the
binding energy to around 8 MeV and a further change of $M_\pi$ to 700 MeV
(corresponding roughly to the HAL QCD calculation \cite{Inoue}) yields then 13 MeV. 
 
Note that the dependence on $M_\pi$ obtained agrees -- at least 
on a qualitative level -- with that presented in Ref.~\cite{Beane11}. 
Specifically, our calculation exhibits the same trend (a decrease of the binding 
energy with decreasing pion mass) and our binding energy of 9 MeV at the physical 
pion mass is within the error bars of the results given in~\cite{Beane11}. 
On the other hand, we clearly observe a non-linear dependence of the binding
energy on the pion mass. As a consequence, scaling our results to the binding
energy reported by the HAL QCD Collaboration \cite{Inoue} (30-40 MeV for  
$M_\pi \approx 700-1000$~MeV) yields binding energies of more than 20~MeV
at the physical point, which is certainly outside of the range suggested in
Ref.~\cite{Beane11}. However, it has to be said that for such large pion masses 
the LO chiral EFT can not be trusted quantitatively. 

Now let me look at the dependence of the $H$ binding energy on the masses of the involved
baryons. In case of the $H$-dibaryon one is dealing with three coupled channels, 
namely $\La\La$, $\Xi N$, and $\Si\Si$. 
Since we know from our experience with coupled-channel problems 
\cite{Polinder:2006zh,Hai10a,Hai05,Hai11} that coupling effects are sizeable and 
the actual separation of the various thresholds plays a crucial role, we
expect a considerable dependence of the $H$ binding energy on the thresholds 
(i.e. on the $\Sigma$, and on the $\Xi$ and $N$ masses). 
Corresponding results are displayed in Fig.~\ref{fig:mpi} in the right panel. 
Note that the pion mass and the masses of the other pseudo-scalar mesons are 
kept at their physical value while varying the $BB$ thresholds. 
For the isospin-averaged masses used in the actual calculation 
the thresholds are at 2231.2, 2257.7, and 2385.0~MeV, respectively. 
Thus, the physical difference between the $\La\La$ and $\Xi N$ thresholds is around
26~MeV while the $\Sigma \Sigma$ threshold is separated from the one for $\La\La$
by roughly 154 MeV.

First I discuss the effect of the $\Sigma \Sigma$ channel because its 
threshold is quite far from the one of $\La\La$ so that there is a rather
drastic breaking of the SU(3) symmetry. Indeed,
when the $\Sigma$ mass is decreased so that the nominal $\Sigma \Sigma$ threshold 
(at 2385~MeV) moves downwards and finally coincides with the one of the 
$\La\La$ channel (2231.2~MeV), a concurrent fairly drastic increase in the $H$ binding 
energy is observed, cf. the solid curve in Fig.~\ref{fig:mpi} for
results based on the interaction with a binding energy of 1.87~MeV for 
physical masses of the mesons and baryons. 
In this context I want to point out that the direct
interaction in the $\Sigma\Sigma$ channel is actually repulsive for the low-energy
coefficients fixed from the $YN$ data plus the pseudoscalar
meson exchange contributions with coupling constants determined from the SU(3)
relations \cite{Polinder:2006zh}, and it remains repulsive even for LEC values that 
produce a bound $H$-dibaryon. But the coupling between the channels generates a sizeable
effective attraction which increases when the channel thresholds come closer. 
The dashed curve is a calculation with the contact term fixed to simulate 
the binding energy ($13.2$~MeV) of the NPLQCD Collaboration at $M_\pi=389\,$MeV.
As one can see, the dependence of the binding energy on the $\Sigma$ mass is rather
similar. The curve is simply shifted downwards by around 4.5~MeV, i.e. by the difference
in the binding energy observed already at the physical masses. 
The asterisks and crosses represent results where, besides the variation of the 
$\Sigma\Sigma$ threshold, the $\Xi N$ threshold is shifted to coincide with that
of the $\La\La$ channel. This produces an additional increase of the $H$ binding 
energy by 20~MeV at the physical $\Sigma\Sigma$ threshold and by 9~MeV 
for that case where all three $BB$ threshold coincide. 
Altogether there is an increase in the binding energy of roughly 60~MeV 
when going from the physical point to the case of baryons with identical
masses, i.e. to the SU(3) symmetric situation. This is significantly larger than 
the variations due to the pion mass considered before. 

\begin{figure}
\centering
\includegraphics[height=6.cm,keepaspectratio,angle=-90]{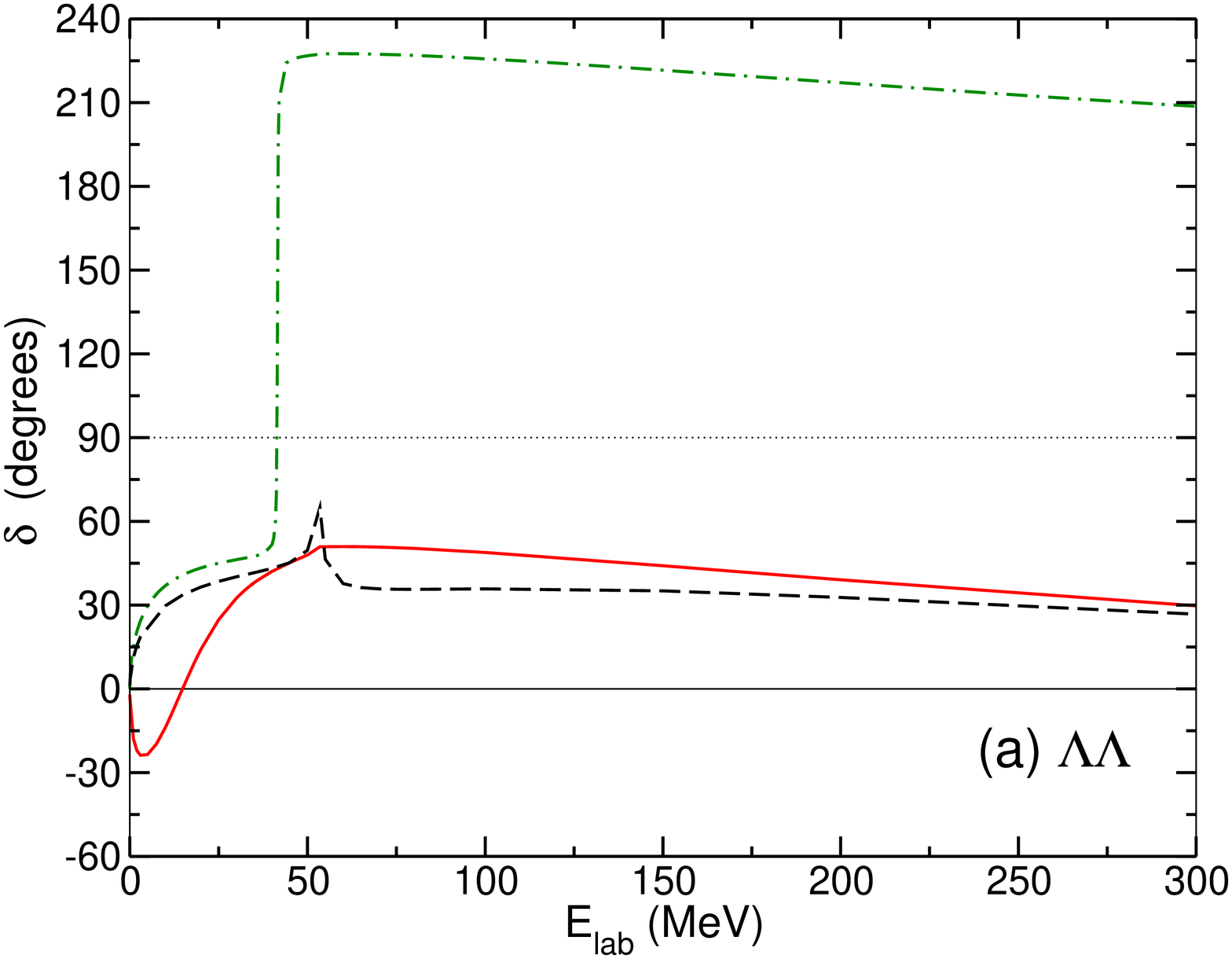}
\includegraphics[height=6.cm,keepaspectratio,angle=-90]{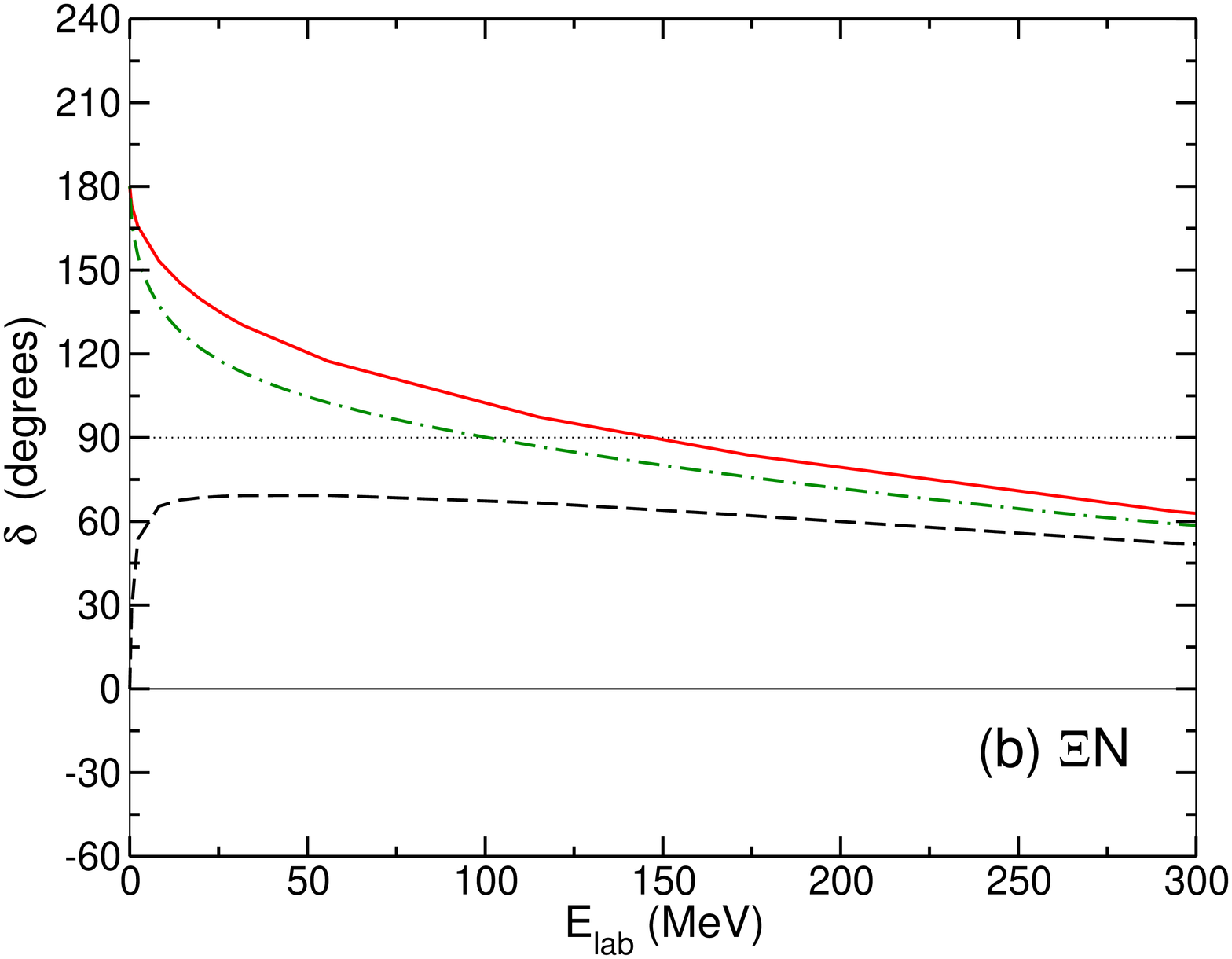}
\caption{Phase shifts in the $^1S_0$ partial wave in the $I=0$ channel of $\La\La$ (a) 
and $\Xi N$ (b) as a function of the pertinent laboratory energies.
The solid line is the result for our illustrative $BB$ interaction that produces a bound $H$ at
$E_H=-1.87\,$MeV.
The other curves are results for interactions that are fine-tuned to the
$H$ binding energies found in the lattice QCD calculations of the HAL QCD (dashed) and
NPLQCD (dash-dotted) Collaborations, respectively, for the pertinent meson (pion) and
baryon masses as described in the text.
}
\label{fig:phases}
\end{figure}

After these exemplary studies let me now try to connect with the published 
$H$ binding energies from the lattice QCD calculations \cite{Inoue,Beane11a}. 
The results obtained by the HAL QCD Collaboration are obviously for the SU(3)
symmetric case and the corresponding masses are given in Table I of 
Ref.~\cite{Inoue}. Thus, one can take those masses and then fix the additional 
LEC so that their $H$ binding energy is reproduced with those masses. 
To be concrete: the masses $M_{P} = 673\,$MeV and $m_{B} = 1485\,$MeV are used, and 
the LEC is fixed so that $E_H = -35\,$MeV. When now the masses of the baryons and
mesons are changed towards their physical values the bound state moves up to the
$\La\La$ threshold, crosses the threshold, crosses also the $\Xi N$ threshold and 
then disappears. In fact, qualitatively this outcome
can be already read off from the curves in Fig.~\ref{fig:mpi} by combining the
effects from the variations in the pion and the baryon masses. Based on those
results one expects a shift of the $H$ binding energy in the order of
60 to 70~MeV for the mass parameters of the HAL QCD calculation. 

In case of the NPLQCD calculation the values provided in 
Ref.~\cite{Beane11b} are taken. Those yield
then 17~MeV for the $\Xi N$-$\La\La$ threshold separation 
(to be compared with the physical value of roughly 26~MeV) and 
77~MeV for the $\Si\Si$-$\La\La$ separation (physical value around 154~MeV). 
We also use the meson masses of Ref.~\cite{Beane11b}, specifically
$M_\pi = 389$ MeV. 
With those baryon and meson masses again the LEC is fixed so that the 
$H$ binding energy given by the NPLQCD Collaboration is reproduced, namely 
$E_H =-13.2$~MeV \cite{Beane11a}.
Again the masses of the baryons and mesons are changed so that they approach 
their physical values. Also here the bound state moves up to and crosses
the $\La\La$ threshold. However, in the NPLQCD case the state survives
and remains below the $\Xi N$ threshold at the physical point. 
Specifically, an unstable bound state \cite{Kok} is observed in 
the $\Xi N$ system around 5~MeV below its threshold and a corresponding 
resonance at a kinetic energy of 21~MeV in the $\La\La$ system 

Phase shifts for the $\La\La$ and $\Xi N$ channels are presented in 
Fig.~\ref{fig:phases}, for the relevant partial wave ($^1S_0$).
The solid line is the result for the $BB$ 
interaction that produces a loosely bound $H$-dibaryon with $E_H=-1.87$~MeV.
The phase shift for the $\Xi N$ channel, Fig.~\ref{fig:phases} (b), is rather 
similar to the one for the $^3S_1$ $NN$ partial wave where the deuteron is
found, see e.g. \cite{Epe05}. 
Specifically, it starts at $180^\circ$, decreases smoothly and eventually 
approaches zero (for large energies not shown in the figure). 
The result for $\La\La$, Fig.~\ref{fig:phases} (a), behaves rather differently. 
The pertinent phase commences at zero degrees, is first negative but becomes positive 
within 20~MeV and finally turns to zero again for large energies.
This behaviour of the phase shifts was interpreted in \cite{Haidenbauer11} 
as a signature for that the bound $H$-dibaryon is actually predominantly a (bound) 
$\Xi N$ state. Indeed, in that work it was argued that it follows already
from the assumed (approximate) SU(3) symmetry of the interaction, that
any $H$-dibaryon is very likely a bound $\Xi N$ state rather than a $\La\La$
state. 
 
The dashed curve corresponds to the interaction that was fitted to the result 
of the HAL QCD Collaboration and reproduces their bound $H$-dibaryon with
their meson and baryon masses. The results in Fig.~\ref{fig:phases} are those
obtained with physical masses of the mesons and baryons. 
The phase shift of the $\Xi N$ channel shows no trace of a bound state anymore. 
Still the phase shift rises up to around $60^\circ$ near threshold, a behavior 
quite similar to that of 
the $^1S_0$ $NN$ partial wave where there is a virtual state (also called
antibound state \cite{Kok}). Indeed, such a virtual state also seems to 
be present in the $\Xi N$ channel as a remnant of the original bound state.
The effect of this virtual state can be seen in the $\La\La$ phase shift where it
leads to an impressive cusp at the opening of the $\Xi N$ channel, 
cf. the dashed line in Fig.~\ref{fig:phases} (a). 

The $\Xi N$ phase shift for the NPLQCD scenario (i.e. for the interaction that 
reproduces their bound $H$-dibaryon with their meson and baryon masses),
see the dash-dotted curve, starts at $180^\circ$, a clear indication for the 
presence of a bound state. However, in this case the bound state is not located 
below the $\La\La$ threshold but above, as already mentioned before. 
Consequently, the corresponding $\La\La$ phase shift exhibits a resonance-like 
behavior at the energy where the (now quasi) bound $H$-dibaryon is located.  

Phase shifts for the $^1S_0$ $\Si\Si$ partial wave can be found in 
Ref.~\cite{Haidenbauer12}. The predictions of the three considered cases
for this channel are practically the same.

\section{Summary}
\label{chap:5}

Chiral effective field theory, successfully applied in Ref.~\cite{Epe05} 
to the $NN$ interaction, also works well for the baryon-baryon interactions 
in the strangeness $S=-1$ ($\Lambda N - \Sigma N$) and $S=-2$ 
($\La \La -\Xi N - \Si\Si$) sectors. 
As shown in our earlier work, already at leading order the bulk properties of 
the $\Lambda N$ and $\Sigma N$ systems can be reasonably well accounted for. 
The new results for the $YN$ interaction presented here, obtained to 
next-to-leading order in the Weinberg counting, look very promising. 
First there is a visible improvement in the quantitative reproduction of
the available data on $\La N$ and $\Si N$ scattering and, secondly, the 
dependence on the regularization scheme is strongly reduced as compared to
the LO result. Indeed the 
description of the $YN$ system achieved at NLO is now on the same level of 
quality as the one by the most advanced meson-exchange $YN$ interactions. 

The recently reported evidence for the so-called $H$-dibaryon  
from lattice QCD calculations stimulated us to investigate also the quark-mass 
dependence of binding energies for baryon-baryon systems in the strangeness $S=-2$ 
sector within the chiral EFT framework. Here I presented results of an analysis 
performed at leading order in the Weinberg counting. 
We found rather drastic effects caused by the SU(3) breaking related 
to the values of the three thresholds $\Lambda\Lambda$, $\Sigma\Sigma$ and $\Xi N$.
For physical values the binding energy of the $H$ is reduced by as much as 60~MeV 
as compared to a calculation based on degenerate (i.e. SU(3) symmetric) $BB$ 
thresholds. 
Translating this observation to the lattice QCD results reported by 
the HAL QCD Collaboration \cite{Inoue},
we see that the bound state has disappeared at the physical point. 
For the case of the NPLQCD calculation \cite{Beane11a}, 
a resonance in the $\Lambda\Lambda$ system might survive.

\section{Acknowledgements}
I would like to thank N. Kaiser, U.-G. Mei{\ss}ner, A. Nogga, S. Petschauer, 
and W. Weise for collaborating on the topic covered by my talk. 
Work supported in part by DFG and NSFC (CRC 110).


\begin{thebibliography}{00}

\bibitem{Wei90}
S.~Weinberg, Phys. Lett. B {\bf 251} (1990) 288.

\bibitem{Wei91}
S.~Weinberg, Nucl. Phys. B {\bf 363} (1991) 3.

\bibitem{Entem:2003ft}
D.~R. Entem, R.~Machleidt, Phys. Rev. C {\bf 68} (2003) 041001.

\bibitem{Epe05}
  E.~Epelbaum, W.~Gl\"ockle, U.-G.~Mei{\ss}ner,
  Nucl.\ Phys.\ A {\bf 747} (2005) 362.

\bibitem{Bed02}
P.~F. Bedaque, U.~van Kolck, Annu. Rev. Nucl. Part. Sci. {\bf 52} (2002) 339.

\bibitem{Epelbaum:2005pn}
  E.~Epelbaum,
  Prog.\ Part.\ Nucl.\ Phys.\ {\bf 57} (2006) 654.

\bibitem{Epelbaum:2008ga} 
  E.~Epelbaum, H.~-W.~Hammer and U.-G.~Mei{\ss}ner,
  Rev.\ Mod.\ Phys.\  {\bf 81} (2009) 1773.

\bibitem{Polinder:2006zh}
  H.~Polinder, J.~Haidenbauer and U.-G.~Mei{\ss}ner,
  Nucl.\ Phys.\ A {\bf 779} (2006) 244.

\bibitem{Polinder:2007mp}
  H.~Polinder, J.~Haidenbauer and U.-G.~Mei{\ss}ner,
  Phys.\ Lett.\ B {\bf 653} (2007) 29.

\bibitem{Hai10a}
  J.~Haidenbauer, U.-G.~Mei{\ss}ner,
  Phys.\ Lett.\ {\bf B684} (2010) 275.

\bibitem{YNNL1} S. Petschauer, diploma thesis, TU Munich, 2011. 

\bibitem{YNNL2} J. Haidenbauer et al., in preparation. 

\bibitem{Petschauer} S. Petschauer, these proceedings. 

\bibitem{Haidenbauer:2007ra}
  J.~Haidenbauer, U.-G.~Mei{\ss}ner, A.~Nogga and H.~Polinder,
  Lect.\ Notes Phys.\ {\bf 724} (2007) 113.

\bibitem{VP}
  C.M. Vincent and S.C. Phatak,          
  Phys.\ Rev.\ C {\bf 10} (1974) 391.

\bibitem{Hai05}
  J.~Haidenbauer, U.-G.~Mei{\ss}ner,
  Phys.\ Rev.\ C {\bf 72} (2005) 044005.

\bibitem{Rij99}
T.~A. Rijken, V.~G.~J. Stoks, Y.~Yamamoto, Phys. Rev. C {\bf 59} (1999) 21.

\bibitem{Nog12} A.~Nogga, these proceedings. 

\bibitem{Haidenbauer11}
  J.~Haidenbauer, U.-G.~Mei{\ss}ner,
  Phys.\ Lett.\ B {\bf 706} (2011) 100.

\bibitem{Haidenbauer12} 
  J.~Haidenbauer and U.-G.~Mei{\ss}ner,
  Nucl.\ Phys.\ A {\bf 881} (2012) 44.

\bibitem{Ahn:2005jz}
  J.~K.~Ahn {\it et al.},
  Phys.\ Lett.\  B {\bf 633} (2006) 214.

\bibitem{Takahashi:2001nm}
  H.~Takahashi {\it et al.},
  Phys.\ Rev.\ Lett.\  {\bf 87} (2001) 212502.

\bibitem{Gal}
  I.~N.~Filikhin, A.~Gal, V.~M.~Suslov,
  Phys.\ Rev.\ C {\bf 68} (2003) 024002.

\bibitem{Rijken}
  T.~A.~Rijken, Y.~Yamamoto,
  Phys.\ Rev.\ C {\bf 73} (2006) 044008.

\bibitem{Fujiwara}
  Y.~Fujiwara, Y.~Suzuki, C.~Nakamoto,
  Prog.\ Part.\ Nucl.\ Phys.\  {\bf 58} (2007) 439.

\bibitem{Yoon}
C.J. Yoon {\it et al.}, Phys. Rev. C {\bf 75} (2007) 022201. 

\bibitem{Ashot}
  A.M. Gasparyan, J. Haidenbauer, C. Hanhart, 
  Phys.\ Rev.\ C {\bf 85} (2012) 015204. 

\bibitem{Jaffe:1976yi}
R. L. Jaffe, Phys. Rev. Lett. {\bf 38} (1977) 195 [Erratum-ibid. {\bf 38} (1977) 617]. 

\bibitem{Beane}
  S.~R.~Beane {\it et al.},
  Phys.\ Rev.\ Lett.\  {\bf 106} (2011) 162001.
\bibitem{Inoue}
  T.~Inoue {\it et al.},, 
  Phys.\ Rev.\ Lett.\  {\bf 106} (2011) 162002.
\bibitem{Beane11a}
  S.~R.~Beane {\it et al.},
  Phys.\ Rev.\ D {\bf 85} (2012) 054511.
\bibitem{Inoue11a}
  T.~Inoue {\it et al.},
  Nucl.\ Phys.\ A {\bf 881} (2012) 28.

\bibitem{Beane11}
 S.~R.~Beane {\it et al.},
  Mod. Phys. Lett. A {\bf 26} (2011) 2587. 
\bibitem{Shanahan11}
   P.~E.~Shanahan, A.~W.~Thomas, R.~D.~Young, 
   Phys.\ Rev.\ Lett.\  {\bf 107} (2011) 092004.

\bibitem{Hai11}
  J.~Haidenbauer, G.~Krein, U.-G.~Mei{\ss}ner, L.~Tolos,
  Eur.\ Phys.\ J.\ A {\bf 47} (2011) 18. 
\bibitem{Beane11b}
  S.~R.~Beane et al., 
  Phys.\ Rev.\ D {\bf 84} (2011) 014507.

\bibitem{Kok} A.~M.~Badalyan, L.~P.~Kok, M.~I.~Polikarpov, Yu.~A.~Simonov, 
  Phys. Rept. {\bf 82} (1982) 31. 

\end{thebibliography}
\end{document}